\documentclass[nofootinbib,prd,floatfix,preprintnumbers,amsmath,amssymb,groupedaddress,superscriptaddress]{revtex4}
\usepackage{dsfont}
\usepackage{epsfig}
\newcommand{\lag}{\mathcal{L}}

\newcommand{\GeV}{\,\mathrm{GeV}}

\newcommand{\tr}{\,\mathrm{Tr}}

\newcommand{\redh}{{ h}}
\newcommand{\redeta}{{\eta}}
\newcommand{\vev}[1]{\langle {#1} \rangle}
\newcommand{\pslash}{\!\not\! p}

\begin{document}
\title{Beyond the Minimal Composite Higgs Model}
\author{Ben Gripaios}
\email{gripaios@cern.ch}
\affiliation{CERN PH-TH, Geneva 23, 1211 Switzerland}

\author{Alex Pomarol}
\email{alex.pomarol@uab.cat}
\affiliation{Departament  de F\'isica, Universitat Aut{\`o}noma de Barcelona,
08193 Bellaterra, Barcelona}

\author{Francesco Riva}
\email{francesco.riva@cern.ch}
\affiliation{CERN PH-TH, Geneva 23, 1211 Switzerland}
\affiliation{D\'epartement de Physique Th\'eorique, Universit\'e de Gen\`eve,
24, Quai Ernest Ansermet, Geneva 4, 1211 Switzerland}

\author{Javi Serra}
\email{jserra@ifae.es}
\affiliation{Departament  de F\'isica, Universitat Aut{\`o}noma de Barcelona,
08193 Bellaterra, Barcelona}

\begin{abstract}
The Higgs spectrum
of the minimal composite Higgs model, based on the 
$SO(5)/SO(4)$ coset,
consists of a unique Higgs doublet whose  phenomenology
does not differ greatly from the Standard Model (SM).
Nevertheless,   extensions   beyond this minimal coset structure
exhibit a richer Higgs spectrum and therefore very different Higgs physics.
We explore one of these  extensions, the $SO(6)/SO(5)$ model,
whose Higgs spectrum contains a $CP$-odd singlet scalar, $\eta$, in addition to the Higgs doublet.
Due to the
pseudo-Nambu-Goldstone
nature of these Higgs bosons,   their physical properties  can be derived from symmetry
considerations alone.
We find that the mass of $\eta$ can be naturally  light,  opening up the possibility that the SM Higgs decays predominantly to the singlet, and therefore lowering the LEP bound on its mass to 86 GeV.
We also show that $\eta$ can have interesting consequences in flavour-violating
processes,  as well as induce  spontaneous  $CP$-violation in the Higgs sector.
The model can also have  anomalies, giving rise to  interactions  between the SM gauge bosons and $\eta$ which,
if  measured at the LHC,   would give quantitative information about the structure of the high energy theory.
\end{abstract}
\maketitle
\section{Introduction}
Models of electroweak symmetry breaking (EWSB) with composite Higgs bosons \cite{Kaplan:1983fs,Georgi:1984af,Dugan:1984hq} have been recently reconsidered \cite{Agashe:2004rs,Contino:2006qr,Barbieri:2007bh,Medina:2007hz,Panico:2008bx,Hosotani:2008tx,Anastasiou:2009rv},  following the stimulus of the AdS/CFT correspondence. In such models, the electroweak scale, $\Lambda_S\sim $ TeV, arises via strong-coupling effects
(just as in QCD the GeV scale arises from the QCD coupling becoming strong), while   the Higgs scalars appear  as pseudo-Nambu-Goldstone bosons (PNGBs) of an approximate symmetry that is non-linearly realized at the electroweak scale.

The presence of strong coupling means that we are powerless to compute in general (at least in situations where the crutch of AdS/CFT is unavailable). Nevertheless, the low-energy
physics of the PNGB Higgs   can be described by an effective lagrangian whose terms
are determined by symmetry considerations, allowing us to study them
without detailed knowledge of the strong sector.
This situation is similar to the pions in QCD, which at low-energies  can be described by the chiral lagrangian based on the symmetry breaking pattern $SU(2)_L \times SU(2)_R\rightarrow SU(2)_{L+R}$.

In the case of models of EWSB,  we have not yet made enough observations  to fully determine what  the symmetry breaking structure, $G\rightarrow H$, is.
The only requirements for the symmetry pattern  in the strong sector are {\em (i)}  $G$ must contain the SM gauge group,  {\em (ii)}  the PNGBs parametrizing  the coset $G/H$  must  contain a Higgs  doublet, and  {\em (iii)}  $H$ must contain a custodial $O(4)$-symmetry to protect
$\Delta\rho$  (or the $T$-parameter) \cite{Sikivie:1980hm}
and $Z\rightarrow b \overline{b}$  \cite{Agashe:2006at} from sizable corrections.

The minimal model  fulfilling the properties  {\em (i-iii)}
is the $SO(5)/SO(4)$ model
\cite{Agashe:2004rs,Contino:2006qr}, whose
sigma model effective lagrangian
contains $4$ NGBs,
making up a complex Higgs $SU(2)_L$-doublet.
The $SO(5)$ symmetry is broken by couplings to SM gauge bosons and fermions, such that
these NGBs become  PNGBs,  getting a potential at the loop-level
and driving EWSB.
The measured value of the $S$-parameter is the only nuisance,
but it appears that this too can be accommodated if one is willing to accept a tuning in the model parameters at a level of no more than one part in ten \cite{Agashe:2004rs,Contino:2006qr}.

Given that the $SO(5)/SO(4)$ model provides a reasonable explanation of existing data, is there any reason to explore less minimal models with an enlarged Higgs sector?
One motivation is that, as stressed above, we do not yet know what the symmetry structure is. The LHC will hopefully settle this question, but in order that it may do so,
we need to be able to identify the different LHC signatures of models with different symmetry structures.
As we shall see, in less minimal models the phenomenology can be dramatically changed, with implications for Higgs physics, flavour physics, and $CP$.
In particular, a new Higgs decay channel can allow
the lower bound of 114 GeV on the value of the SM Higgs mass to be evaded, and
can accommodate a lighter Higgs, as preferred in composite scenarios.

Another motivation is that, as we will learn in Section~\ref{wzw},  in less minimal models with a different symmetry structure, we have the possibility of non-trivial physics
associated with quantum anomalies of the symmetry.
Since the anomaly is non-renormalized, the coefficients of these operators are completely fixed, up to integers that measure the fermion content of the high-energy theory. If we were able to measure these integers at the LHC or a future collider, we would be able to obtain quantitative information about the ultra-violet (UV) theory, similarly to the way in which the decay
$\pi^0\rightarrow\gamma\gamma$ allowed us to extract the number of colours in QCD.

In this Article, we explore these issues in  one of the simplest extensions of the minimal composite Higgs model, the model  based on the coset $SO(6)/SO(5)$.\footnote{This coset was previously explored in the context of little Higgs models in Ref.~\cite{Katz:2005au}, and also in Ref.~\cite{Lodone:2008yy}.}
The model contains  $15-10=5$ NGBs, comprising a SM Higgs doublet and an electroweak singlet $\eta$.
The presence of  $\eta$ can lead to   interesting
and varied implications for phenomenology
that, as we will see,   crucially depend on the embedding of   the SM fermions into representations of the global $SO(6)$.
We will see that these  embeddings     can
preserve the symmetry associated with shifts of the NGB $\eta$,  and protect
the $\eta$ mass  from  SM loop corrections.
In particular,  we will present  a scenario in which
the gauge and the top contributions to the $\eta$ mass
are zero, and therefore $\eta$ can be naturally light, $\lesssim 30$ GeV, getting
its mass  from bottom or tau loops.
This opens up the possibility of decays of the SM Higgs into the singlet, invalidating the LEP bound on the Higgs mass.
The dominant decay channel of $\eta$ can be $b\bar b$, $\tau\bar\tau$ or $c\bar c$,
depending on  the corresponding embeddings into $SO(6)$ of  the remaining SM fermions.
If  the embeddings are  different for different
family members,
we will show  that  the $\eta$ can mediate flavor-changing neutral currents (FCNC)
and have flavour-violating decays.
Furthermore, the  model incorporates extra sources of  $CP$-violation,
with important implications in the Higgs sector.
Since the group $SO(6)$ is isomorphic to $SU(4)$, the model  can have an anomaly, and correspondingly a Wess-Zumino-Witten (WZW) term. This  term generates a coupling between $\eta$  and two SM gauge bosons, and  could be measured
by detecting the decay channel $\eta\rightarrow\gamma\gamma$.

We will also explore models based on the $SO(6)/SO(4)$ coset
containing  two Higgs doublets.
Nevertheless, we will show that in these models the custodial symmetry
is generically broken, implying that contributions to the $T$-parameter are large.

The layout is as follows. In Section~\ref{nmchm} we introduce the $SO(6)/SO(5)$ model,
describing  how the SM fields are coupled to the Higgs.
This allows us to determine the form of  the Higgs potential, and discuss  the resulting phenomenology. In Section~\ref{wzw}
we provide a discussion of anomalies and the WZW term in models based on general cosets. We give a necessary condition for a WZW term to arise, and show that this is fulfilled  in the
$SO(6)/SO(5)$ model.
We conclude in Section~\ref{conclu}.
In Appendix~\ref{4}, we consider a similar model based on $SO(6)/SO(4)$
and show that it generically  does not preserve  the  custodial symmetry.
Appendix~\ref{CP} discusses $C$ and $P$
in the Higgs sector of the
$SO(6)/SO(5)$ model, in the presence of a WZW term.
\section{The $SO(6)/SO(5)$ composite Higgs model}\label{nmchm}

In the case that the global symmetry breaking of the strong sector is
$SO(6) \cong SU(4)\rightarrow SO(5)\cong Sp(4)$ the model will contain
five  NGBs, transforming  as  a ${\bf 5}$ of $SO(5)$, which corresponds to a
$\mathbf{1} \oplus \mathbf{4} \equiv (\mathbf{1},\mathbf{1}) \oplus (\mathbf{2},\mathbf{2})$
under  the subgroup $SO(4) \cong SU(2)_L \times SU(2)_R$.
The bi-doublet  can be associated  to the usual SM  Higgs doublet $H$
responsible for EWSB,
while the  singlet, which we denote by $\eta$, corresponds to an extra pseudoscalar state.
The breaking of  $SU(4)$ down to $Sp(4)$ can be achieved  by
a  $4\times4$ antisymmetric matrix
\begin{gather}\label{SigmaVEV}
\Sigma_0=
\begin{pmatrix}
i\sigma_2  &0    \\
0 &     i\sigma_2
\end{pmatrix},
\end{gather}
corresponding to the vacuum expectation value (VEV) of a field $\Sigma$ transforming as the \textbf{6}
of $SU(4)$:
\begin{gather}\label{SigmaTrafo}
\Sigma\rightarrow U\Sigma U^T\, .
\end{gather}
The  unbroken generators  $T^a$ satisfy
\begin{gather}\label{unbroken}
T^a \Sigma_0 + \Sigma_0 T^{aT} = 0\, ,
\end{gather}
and correspond to  the generators of $Sp(4)\cong SO(5)$,  while
the broken ones, $T^{\hat{a}}$,  satisfy
\begin{gather}\label{broken}
T^{\hat{a}} \Sigma_0 - \Sigma_0 T^{\hat{a}T} = 0 \, .
\end{gather}
Among the ten unbroken generators we identify
six corresponding to the subgroup $SU(2)_L\times SU(2)_R$ as
\begin{gather}
T^{{a}}_L = \frac{1}{2}\begin{pmatrix}  \sigma_a & 0\\  0 &  0 \end{pmatrix}, \, \, T^{{a}}_R =  \frac{1}{2}\begin{pmatrix}  0 & 0\\  0 &  \sigma_a \end{pmatrix},
\end{gather}
while the remaining  four can be taken to be
\begin{gather}
\frac{1}{2\sqrt{2}}\begin{pmatrix}  0 & \sigma_a \\  \sigma_a\ &  0 \end{pmatrix} \, \, \mathrm{and} \, \, \,\frac{1}{2 \sqrt{2}} \begin{pmatrix}  0 & -i\mathds{1} \\  +i\mathds{1} &  0 \end{pmatrix}.
\end{gather}
The fluctuations along the broken directions correspond to the NGBs, which parametrize
the $SU(4)/Sp(4)$ coset
\begin{gather}
\label{SigmaPi}
\Sigma=e^{\frac{i}{\sqrt{2}}\Pi/f}\Sigma_0\, ,
\end{gather}
where
\begin{gather}
\Pi = \begin{pmatrix} \eta \mathds{1} & -i (H^{c} \; H) \\ i( H^{c} \;H)^\dagger & -\eta \mathds{1} \end{pmatrix},
\end{gather}
with $H=\begin{pmatrix}h_3+ih_4\\ h_1+ih_2\end{pmatrix}$  and $H^c=i\sigma_2 H^*$.
This can be written as
\begin{gather}\label{Sigma}
\Sigma  =  \begin{pmatrix} \left(c + i \frac{\eta s}{\sqrt{\eta^2 + h^2}} \right)i\sigma_2 & \frac{s}{\sqrt{\eta^2 + h^2}} (-H \; H^c) \\  - \frac{s}{\sqrt{\eta^2 + h^2}} (-H\; H^c)^T &
\left(c- i \frac{\eta s}{\sqrt{\eta^2 + h^2}} \right)i\sigma_2  \end{pmatrix},
\end{gather}
where
\begin{equation}
s=\sin\frac{\sqrt{\eta^2+h^2}}{\sqrt{2}f}\, , \ \ \  c=\cos\frac{\sqrt{\eta^2+h^2}}{\sqrt{2}f}\, , \ \ {\rm and} \ \ h=\sqrt{h_i^2}\, .
\end{equation}
By a suitable $SU(2)_L$ rotation, we can eliminate 3 NGBs (they are eaten by the SM gauge bosons), and keep only the physical Higgs, $h$, and $\eta$.
In this gauge, the kinetic term for the PNGBs is given by
\begin{equation}
\label{kin}
\frac{f^2}{8}\tr |D_{\mu} \Sigma|^2 = \frac{f^2}{2} (\partial_\mu h)^2+
\frac{f^2}{2} (\partial_\mu\eta)^2
+\frac{f^2}{2}\frac{(h\partial_\mu h+\eta\partial_\mu\eta)^2}{1-h^2-\eta^2}
+\frac{g^2f^2}{4}h^2\left[W^{\mu+} W^-_\mu+\frac{1}{2\cos^2\theta_W}Z^\mu Z_\mu\right]\, ,
\end{equation}
where we have performed the following redefinition of the PNGB fields:
\begin{equation}
\frac{h^2 s^2}{\eta^2 + h^2}\rightarrow h^2\, , \ \ \ \
\frac{\eta^2 s^2}{\eta^2 + h^2}\rightarrow \eta^2\, .
\label{def}
\end{equation}
Field choices related by re-definitions of this type are equally valid inasmuch as the sigma-model itself is concerned \cite{Coleman:1969sm}, but, as is clear from Eq.~(\ref{kin}), the redefined $h$ is the one whose VEV sets the scale of EWSB.
From now on, $h$ and $\eta$ will always refer to the redefined  fields.

The gauging of the SM group breaks the global symmetry~\footnote{In general, gauging a subgroup $K$ of a global symmetry breaks the global symmetry down to the largest  subgroup that contains $K$ as an ideal.} $SU(4)$ down to
$SU(2)_L \times U(1)_Y\times U(1)_{\eta}$, where $Y=T_R^3$ and
$U(1)_{\eta}$ is generated by
\begin{equation}
\label{teta}
T^\eta=\frac{1}{2\sqrt{2}}{\rm Diag}(1,1,-1,-1)\, .
\end{equation}
Since this latter is the symmetry under which the PNGB $\eta$ shifts,   gauge boson loops  will generate a potential for $h$, but not for $\eta$.

\subsection{Couplings to SM fermions}
We now consider the couplings of  the strong sector  to the SM fermions.
As in Ref.~\cite{Agashe:2004rs}, we will assume that the SM fermions couple linearly to a single operator
of the  strong sector  (or, equivalently, to a resonance
of the strong sector); these mixings will be the origin of the fermion masses.
For this purpose, we need to enlarge
the global
group of the strong sector to include the colour  group $SU(3)_c$, and an extra
$U(1)_X$, which allows us to properly embed the hypercharges, as $Y = T^3_R + X$.
This extra $U(1)_X$ will not be  spontaneously broken, and therefore its inclusion does not affect the results of the previous section. The PNGB fields have vanishing $X$-charge.

Choosing the quantum numbers  of the operators in the strong sector, to which the SM fermions are coupled, is equivalent  to choosing an embedding  for the SM fermions into representations of the global  $SU(4)\times U(1)_X$.
Since it is not possible to embed  the SM fermions into complete representations,
the couplings between the SM fermions  and the strong sector  will, in general, break the global symmetries.
We will, however, demand  that these couplings preserve the custodial symmetry
that protects $Zb\bar{b}$ from large corrections \cite{Agashe:2006at}. This means that the quark
doublet must be embedded in a ${\bf (2,2)_{2/3}}$ of  $SU(2)_L \times SU(2)_R\times U(1)_X$.
Let us now consider, in turn,
the three smallest representations of $SU(4)$, namely the $\mathbf{4}$, the
$\mathbf{10}$ and the $\mathbf{6}$.\footnote{Similar considerations apply to the conjugate $\mathbf{\overline{4}}$ and $\mathbf{\overline{10}}$ representations.}

The $\mathbf{4}$ decomposes as $(\mathbf{2},\mathbf{1}) \oplus (\mathbf{1},\mathbf{2})$
under $SU(2)_L\times SU(2)_R$, and therefore can be discarded since it does not contain a
$(\mathbf{2},\mathbf{2})$.

The $\mathbf{10}$, a symmetric tensor of $SU(4)$,  decomposes into $(\mathbf{2},\mathbf{2})\oplus(\mathbf{3},\mathbf{1})\oplus (\mathbf{1},\mathbf{3})$ under
$SU(2)_L\times SU(2)_R$.
We can embed the SM quark doublet, $q_L$, into the  $(\mathbf{2},\mathbf{2})$, while
the quark singlets, $u_R$ and $d_R$,  can go into  the $(\mathbf{1},\mathbf{3})$:
\begin{gather}
\Psi_q=\frac{1}{\sqrt{2}}\begin{pmatrix} 0&Q\\ Q^T& 0 \end{pmatrix}, \quad
\Psi_u=\begin{pmatrix} 0 &0 \\ 0 &U\end{pmatrix}, \quad \Psi_d=\begin{pmatrix} 0 & 0 \\ 0 &D\end{pmatrix},
\end{gather}
where
\begin{gather}
Q=\begin{pmatrix} 0& q_L \end{pmatrix}, \quad U=\begin{pmatrix}  0 &u_R \\ u_R & 0 \end{pmatrix}, \quad D=\begin{pmatrix} d_R & 0\\ 0 & 0\end{pmatrix}.
\end{gather}
The $X$-charge  assignments are the following:    $X_q=2/3$, which, as discussed above,  guarantees that the custodial symmetry  protects $Zb\bar{b}$, and
$X_u=X_d=2/3$, in order to allow a Yukawa coupling with $\Sigma$.
We notice, however, that this embedding does not break the global $U(1)_\eta$ symmetry of
Eq.~(\ref{teta}), since $q_L$, $u_R$ and $d_R$ have a
well-defined transformation among themselves. Indeed, under $U(1)_\eta$, we find
\begin{gather}
\label{trans}
\delta \Psi_i = T^\eta \Psi_i + \Psi_i  T^{\eta\, T}\, ,
\end{gather}
whence
\begin{gather}
\delta q_L = 0\,  ,\ \ \
\delta u_R = -\frac{1}{\sqrt{2}} u_R\, , \ \ \
\delta d_R = -\frac{1}{\sqrt{2}} d_R\, .
\end{gather}
That is to say, the SM fermions have well-defined charges under $U(1)_\eta$.
Thus, there is a remnant $U(1)_\eta$ symmetry that is broken neither by gauge nor by Yukawa interactions.
What is more, if this $U(1)_\eta$ is assumed to be  anomalous in the background of QCD,
it will be a  {\em bona fide} Peccei-Quinn symmetry, solving the strong $CP$ problem.
The  $\eta$ will correspond to the  axion, and will  obtain a mass of order $m_\pi f_\pi / f$ via mixing with pions.
Unfortunately, we know that an electroweak-scale axion of this type has been essentially excluded, by searches for $K^+\rightarrow \pi^+ \eta$,
irrespectively of its model-dependent couplings to fermions and gauge bosons
\cite{Georgi:1986df}.
We therefore discard the $\mathbf{10}$ as well.

This leaves us with the last possibility,  namely embedding the SM fermions in the $\mathbf{6}$-dimensional representation of $SU(4)$, carried by antisymmetric $4\times4$ matrices (this is the vector representation of $SO(6)$).
Under $SU(2)_L\times SU(2)_R$, this representation decomposes as $(\mathbf{2},\mathbf{2}) \oplus (\mathbf{1},\mathbf{1}) \oplus (\mathbf{1},\mathbf{1})$; the SM $q_L$ must go into the bi-doublet, while $u_R$ and $d_R$ each go into some linear combination of the two singlets.
For the up-quark sector, we have the embedding
\begin{equation}
\label{embe1}
\Psi_q = \frac{1}{2} \begin{pmatrix}
0    &  Q  \\
-Q^T    & 0
\end{pmatrix}, \quad
\Psi_u =  \Psi_u^+ + \epsilon_u \Psi_u^-\, ,
\quad \Psi_u^{\pm} = \frac{1}{2} \begin{pmatrix}
\pm U   &  0  \\
0   &  U
\end{pmatrix},
\end{equation}
where $Q = (0,q_L)$,  $U =  u_R i \sigma_2$, and the complex parameter $\epsilon_u$
defines the embedding of the $u$-quark into the two singlets.
As in the case of the {\bf 10}, the $X$-charges are  $X_q = +2/3 = X_u$.
For the down-sector,
we  are forced to embed the quark doublet into a second ${\bf 6}$-plet,  $\Psi_{q^\prime}$, with
$X_{q^\prime}=-1/3$.
This is necessary in order to generate non-zero down-type  masses, since
the multiplet containing  the $d$-quark, $\Psi_{d}$, must have $X_d=-1/3$ to give the correct hypercharge to $d_R$.
The embeddings are then given by
\begin{equation}
\label{embe2}
\Psi_{q^\prime} = \frac{1}{2} \begin{pmatrix}
0    &  Q^\prime  \\
-Q^{\prime\, T}    & 0
\end{pmatrix}, \ \ \ \
\Psi_d = \Psi_d^+ + \epsilon_d \Psi_d^-\, ,
\quad \Psi_d^{\pm} = \frac{1}{2} \begin{pmatrix}
\pm D    &  0  \\
0   &  D
\end{pmatrix},
\end{equation}
where now $Q^\prime = (q_L,0)$ and  $D =  d_R i \sigma_2$.
The fact that the $q_L$-doublet  arises from a multiplet with $X=-1/3$ implies that
the custodial symmetry cannot guarantee protection of  the $Z b\bar b$ coupling.
Nevertheless,   this  multiplet can be assumed to be coupled to the strong sector
with a small coupling $\propto \sqrt{m_b}$,  assuring the generation of the bottom mass
without substantially  affecting  the $Z b\bar b$ coupling.
From   Eqs.~(\ref{embe1}) and  (\ref{embe2}) we observe that in the special case
$\epsilon_i =\pm1$ $(i=u,d)$, the SM quarks
have definite charges under   the $U(1)_\eta$ [Eq.~(\ref{trans})]:
\begin{equation}
\delta q_L = 0\, ,\ \ \
\delta u_R =  \mp \frac{1}{\sqrt{2}} u_R\, ,\ \ \
\delta d_R =  \mp \frac{1}{\sqrt{2}} d_R\, .
\end{equation}
Therefore we expect to find a massless $\eta$ in the limit
$\epsilon_i \rightarrow \pm 1$.

\subsection{One-loop effective potential}

At the one-loop level, a potential for the PNGBs is generated due to the $SU(4)$-breaking
terms  arising from  the SM couplings to the strong sector. This potential depends on the
dynamics of the strong sector, which is in general unknown.
Nevertheless, symmetry considerations are powerful enough to tell us  the  functional  form of the potential, and to determine whether  $h$ and $\eta$ can or cannot get a  non-zero VEV, as well
as the size of their masses.
To obtain the functional form of the one-loop effective potential, we proceed in two steps. First, we write the lagrangian for the SM fields obtained by  integrating out the strong sector in the background of $\Sigma$. Second,  we give the one-loop potential
generated by integrating over the SM fields.

Using the invariance
under the global $SU(4)\times U(1)_X$,  we can write the
effective lagrangian for the SM gauge bosons at the  quadratic level, obtained by integrating
out the strong sector,  as
\begin{widetext}
\begin{gather}
\lag_{\rm g}=\frac{1}{2}P^{\mu\nu}\Big(\Pi_0^BB_{\mu}B_{\nu}+\Pi_0\tr\left[A_{\mu}A_{\nu}\right]+\Pi_1\tr\left[(A_{\mu}\Sigma+\Sigma A_{\mu}^T)(A_{\nu}\Sigma+\Sigma A_{\nu}^T)^\dagger\right]\Big)\, ,
\end{gather}
where $B_{\mu}$ is  the $U(1)_Y$ gauge field and  $A_{\mu}=A_{\mu}^{a}T^{a}_L+B_{\mu}T^3_R$ where $A_{\mu}^{a}$  are the gauge fields of $SU(2)_L$.
The lagrangian is given  in momentum-space
and the $\Pi_i$ are momentum-dependent form factors
whose values depend on the strong dynamics. In extra dimensional models these quantities
can be explicitly calculated \cite{Agashe:2004rs}.
We have also defined  $P^{\mu\nu}=\eta^{\mu\nu}-p^{\mu}p^{\nu}/p^2$, where $p$
is the momentum of the gauge fields.
Using Eqs.~(\ref{Sigma}) and (\ref{def}), and the explicit expression for the $SU(2)_L$ generators, we have
\begin{gather}
\label{propgauge}
\lag_{\rm g}=\frac{1}{2}P^{\mu\nu}\left[ \left( \Pi_0^B + \frac{\Pi_0}{2} + \Pi_1 h^2 \right)B_{\mu}B_{\nu}+\left( \frac{\Pi_0}{2} +\Pi_1 h^2\right) A_{\mu}^aA_{\nu}^a - 2\Pi_1 h^2 A_\mu^3 B_\nu \right]\, .
\end{gather}
\end{widetext}
Similarly, for the  SM quarks,  the most general $SU(4)\times U(1)_X$-invariant lagrangian obtained after integrating out the strong sector is given, at the quadratic order, by~\footnote{We have used the fact that $\tr[\bar \Psi_{q} \Psi_{u}] = \tr[\bar \Psi_{q'} \Psi_{d}]=0$ when projected to the SM field content, Eqs.~(\ref{embe1}) and (\ref{embe2}).}
\begin{equation}
\label{strong}
\mathcal{L}_{\rm f} = \sum_{r=q,u,q^\prime,d} \Big[\Pi_0^r \tr[\bar{\Psi}_r \pslash\, \Psi_r] + \Pi_1^r \tr[\bar{\Psi}_r \Sigma]  \pslash  \tr[\Psi_r \Sigma^\dagger]\Big] + M^u_1 \tr[\bar \Psi_q \Sigma] \tr[\Psi_u \Sigma^\dagger] +M^d_1 \tr[\bar\Psi_{q^\prime} \Sigma] \tr[\Psi_d \Sigma^\dagger]+ h.c. \, ,\end{equation}
where we have
\begin{align}
\label{fermionhiggs}
&  \tr[\bar{\Psi}_q \Sigma] \pslash \tr[\Psi_q \Sigma^\dagger] =  \bar u_L  \pslash u_L\, \redh^2\, , \nonumber\\ &\tr[\bar{\Psi}_{q'} \Sigma] \pslash \tr[\Psi_{q'} \Sigma^\dagger] =  \bar d_L  \pslash d_L\, \redh^2\, ,\nonumber\\
& \tr[\bar{\Psi}_u \Sigma] \pslash \tr[\Psi_u \Sigma^\dagger] = 4 \bar u_R \pslash  u_R \left|  \sqrt{1-\redeta^2-\redh^2} + i \epsilon_u \redeta \right|^2 \, ,\nonumber\\
& \tr[\bar{\Psi}_d \Sigma] \pslash \tr[\Psi_d \Sigma^\dagger] = 4 \bar d_R \pslash  d_R \left|  \sqrt{1-\redeta^2-\redh^2} + i \epsilon_d \redeta \right|^2 \, ,\nonumber\\
& \tr[\bar\Psi_q \Sigma] \tr[\Psi_u \Sigma^\dagger] = 2 \bar u_L u_R\, \redh \left[
\sqrt{1-\redeta^2-\redh^2}+ i \epsilon_u \redeta \right]\, ,\nonumber\\
&   \tr[\bar\Psi_{q^\prime} \Sigma] \tr[\Psi_d \Sigma^\dagger] = - 2 \bar d_L d_R\, \redh \left[
\sqrt{1-\redeta^2-\redh^2}+ i \epsilon_d \redeta \right]\, .
\end{align}
The last two terms of Eq.~(\ref{strong}) give rise to the quark masses, so we must require
that at zero momentum $M^{u,d}_1\sim m_{u,d}$.
This can be achieved by requiring that the SM fermion $f_i$ couples to the strong sector
with a strength   $\propto \sqrt{m_{f_i}}$ where $m_{f_i}$ is the fermion mass.\footnote{We are assuming that for a given SM fermion the left-handed and right-handed components have similar  couplings to the strong sector.
This guarantees that all FCNC processes from the strong sector are suppressed  (for a recent analysis see Refs.~\cite{Csaki:2008zd,Blanke:2008zb}).
Relaxing this assumption can lead to  large  FCNC effects.}
This implies
\begin{equation}
\label{propor}
\Pi^q_1, M^u_1\propto m_u\ ,\ \ \ \ \ \
\Pi^{q^\prime}_1, M^d_1\propto m_d\, .
\end{equation}
A similar lagrangian is obtained for the SM leptons.

Now, by integrating out   the SM fields, we can get the effective potential for the PNGBs.
This is expected to be dominated by one-loop effects arising from  the  $SU(2)_L$ gauge bosons  and, due to Eq.~(\ref{propor}),    3rd family quarks.  We find
\begin{equation}
\label{potheta}
V(h,\eta) = \frac{9}{2} \int \frac{d^4 p}{(2\pi)^4} \log{\Pi_W} - (2N_c) \int \frac{d^4 p}{(2\pi)^4} \left[ \log \Pi_{b_L} + \log{(p^2 \Pi_{t_L} \Pi_{t_R} - |\Pi_{t_L t_R}|^2)} \right]\, ,
\end{equation}
where the gauge and top propagators arise respectively from Eqs.~(\ref{propgauge}) and (\ref{strong}) with $u\rightarrow t$:
\begin{align}
& \Pi_W = \frac{\Pi_{0}}{2} + \Pi_1 \redh^2\, ,
\ \ \ \Pi_{t_L} = \frac{\Pi_0^q+\Pi_0^{q^\prime} }{2} - \Pi_1^q  \redh^2\, ,\ \ \ \Pi_{b_L} = \frac{\Pi_0^q+\Pi_0^{q^\prime} }{2} - \Pi_1^{q'}  \redh^2\, , \nonumber  \\
&\Pi_{t_R} = \Pi_0^t - \Pi_1^t 4 \left|  \sqrt{1-\redeta^2-\redh^2} + i \epsilon_t \redeta \right|^2
\, , \ \ |\Pi_{t_L t_R}|^2 = |M^t_1|^2  4 \redh^2
\left| \sqrt{1-\redeta^2-\redh^2} + i \epsilon_t \redeta \right|^2 \, .
\end{align}
The functions $\Pi_1$ and $M_1$ characterize   the effects of the spontaneous
$SU(4)$-breaking  in the strong sector, and therefore
must decrease for momentum  $p$ above the scale of the strong sector $\Lambda_S$.
This allows for  an expansion of the logarithms in the potential that  leads to an  approximate formula for the potential:
\begin{equation}
\label{pot}
V(\redh,\redeta) \simeq \alpha \redh^2+\lambda \redh^4+
|\phi|^2 \left[\beta + \gamma \redh^2
+ \delta |\phi|^2
\right]\, ,  \ \ \ \ \  \phi \equiv \sqrt{1-\redeta^2-\redh^2} + i \epsilon_t \redeta \, ,
\end{equation}
where $\alpha, \lambda, \beta, \gamma,$ and $\delta$ are constants that depend on integrals over the form factors.
In the limit
$\epsilon_t \rightarrow \pm 1$ in which
the $\redeta$ becomes a true NGB, we have
\begin{equation}
\label{phi}
|\phi|^2 \rightarrow (1-\redh^2),
\end{equation}
and therefore the potential Eq.~(\ref{pot}) becomes $\eta$-independent.
In this limit, the potential for $\eta$ may be sensitive to other one-loop effects, coming from
the light SM fermions.
This will be the case for  SM fermions $f_i$ whose embedding parameters $\epsilon_i$   take values different from $\pm 1$.
We shall explore this possibility further later on.

We will be interested in cases in which $h$ gets a VEV and breaks the electroweak symmetry,
with $\eta$ either getting a VEV, or not getting a VEV.
Both of these situations can occur, for suitable values of the parameters.
For example, for complex values of  $\epsilon_t$, $\eta$ gets always a VEV    since the term
$|\phi|^2$ contains  a tadpole for $\eta$.
Notice that, due to the re-definition Eq.~(\ref{def}), the VEVs of the PNGBs must be  restricted to
\begin{equation}
\vev{h^2}+\vev{\eta^2}\leq1\, .
\end{equation}
For $\epsilon_i\in\mathbb{R}$ and   $\vev{\eta}=0$, the Higgs $h$ can be defined as a $CP$-even scalar, while $\eta$ is $CP$-odd,
as can be deduced  from their couplings to fermions in Eq.~(\ref{fermionhiggs}). This
assignment is consistent with the other NGB interactions, as explained in
Appendix~\ref{CP}.
Even if $\epsilon_i\in\mathbb{R}$, we can have  $\vev{\eta}\not=0$, and then $CP$ is spontaneously broken.
We must be aware, however,  that the effects of a nonzero VEV for $\eta$
vanish
in the limit $\epsilon_i \rightarrow \pm 1$, since the $\eta$ becomes a true NGB and therefore its
VEV is unphysical.
When  $\epsilon_i\notin\mathbb{R}$,  $CP$ is explicitly broken
in the $\eta$ interactions to fermions Eq.~(\ref{fermionhiggs}).
In this case we always have, as we explained above, that  $\vev{\eta}\not=0$, and $CP$  is
in fact broken in all Higgs interactions.

\subsection{Higgs phenomenology}
The Higgs physics in this model  strongly depends on the values of $\epsilon_i$
which, without knowledge of  the underlying  strong sector,  must be taken as free parameters.
Two important values for these parameters are
\begin{eqnarray}
\label{specialv}
\epsilon_i=\pm 1&\ \  \implies\ \ & \text{No potential is generated for $\eta$ from loops of the fermion $f_i$}\, ,\nonumber\\
\epsilon_i=0&\ \ \implies\ \ &  \text{Zero  $\eta f_i\bar f_i$ coupling}\, .
\end{eqnarray}
In the following, we discuss different possibilities for $\epsilon_i$ and the phenomenological
implications in Higgs physics.
We first consider  the case in which  $\epsilon_i$  are family  universal,
and consider later the FCNC implications  when this is not the case.

{\bf Heavy-$\eta$ scenario:}
If the value of $\epsilon_t$ is different from $\pm1$, then we have a scenario
in which $\eta$ gets a potential from top-loops. In this case, we
have  two physical Higgs states, $h$ and $\eta$,
with masses around $100-200$ GeV.
If $\vev{\eta}=0$ ($\epsilon_i\in\mathbb{R}$), we have,
up to effects of order $\vev{h^2}$, that $h$ behaves as the SM Higgs.
The  $\eta$  is a $CP$-odd state, and couples to fermions, via  Eq.~(\ref{fermionhiggs}),
with a strength
\begin{equation}
g_{\eta f_if_i}=  m_{f_i}  \frac{\epsilon_i}{\sqrt{1-\langle h^2\rangle}}\, .
\end{equation}
An important difference between $\eta$
and $h$ is the absence of a tree-level coupling  of $\eta$ to
$WW$ and  $ZZ$, {\em cf.} Eq.~(\ref{kin}).
By   measuring,  at the LHC,  the different products  $\sigma\times BR$,  where $\sigma$  represent the different  production rates (either through gluon, gauge-boson fusion, or top-strahlung),
and $BR$   the possible branching ratios (decays into $b$, $\tau$ , $\gamma$   and (virtual) weak gauge bosons), we  can assure the discovery of  the two Higgs states.
Nevertheless, even if these can be measured, the difficult task will be to disentangle this scenario from others, {\em e.g.}, supersymmetric models.
This could be possible if we were able  to obtain a  precise determination at the LHC of the different
values of  the  products  $\sigma\times BR$ that, as explained in Ref.~\cite{Giudice:2007fh},
suffice to establish the composite nature of the Higgs.
Another option would be to try to distinguish  $\eta$   from, for example,  the $CP$-odd scalar $A_0$ of the MSSM. The main difference among the two arises in their coupling to $hZ$, which is present for $A_0$, but absent for $\eta$.
If we can establish the presence of this coupling at the LHC, either from the decay of the $CP$-odd state to the $CP$-even one (or {\em vice versa}), or from the double production of the $CP$-odd and $CP$-even states, this will definitely rule out the scenario considered here.

For the case in which $\vev{\eta}\not=0$, the two Higgs states mix with each other and we end up in a scenario of two Higgs states with very similar phenomenology.
The important implication in this case is that $CP$ is violated in the Higgs sector.
Nevertheless, to observe this we must  rely on the decay of the Higgs to $WW/ZZ$, if kinematically possible, or to $\tau\bar \tau$, whose branching fraction is very small. These are the only two
decay channels that allow a full analysis of the angular distribution of the decay products and
a determination of the $CP$-properties of the Higgs \cite{Accomando:2006ga}.
Another  suggestion is to
use the angular correlations of the tagging jets  in vector boson fusion  production of the Higgs
\cite{Ruwiedel:2007zz}.

{\bf Light-$\eta$ scenario:}
In  the limit in which all $\epsilon_i\rightarrow \pm 1$,
the $\eta$ mass goes to zero, and we are driven to a very different scenario for Higgs physics.
The mass of  $\eta$ can be  below $m_h/2$, implying that the Higgs  $h$
can decay to $\eta\eta$.
From  Eq.~(\ref{kin}) we find  a   $\redh \redeta \redeta$  coupling~\footnote{There is also a coupling in the potential Eq.~(\ref{pot}), but it vanishes as $\epsilon_i\rightarrow \pm 1$.}
\begin{equation} \label{etah}
-\frac{f^2 \vev{\redh}}{2} \redeta^2 \partial^2_\mu \redh \, ,
\end{equation}
which  leads to  a  Higgs partial width
\begin{equation}
\Gamma(\redh\rightarrow
\redeta\redeta)=\frac{ m_\redh^3 m^2_W\beta}{8\pi g^2 f^4}\, , \; \; \beta=\sqrt{1-4 m_\redeta^2/m_\redh^2}\, .
\end{equation}
This decay channel can dominate over the $b\bar b$ channel. In the limit of  $m_\eta\ll m_h$,
we find
\begin{gather}
\frac{\Gamma(\redh\rightarrow \redeta\redeta)}{\Gamma(\redh\rightarrow b\bar b)}
\simeq 8.5\, \Big(\frac{m_h}{120\ {\rm GeV}}\Big)^2\Big(\frac{500\ {\rm GeV}}{f}\Big)^4\, .
\end{gather}
This opens up the possibility  that the Higgs could in fact be somewhat lighter than the LEP SM Higgs bound of 114 GeV, since
$h$  might have escaped detection at LEP  due to the non-standard decay mode
$h\rightarrow \eta\eta$ \cite{Dobrescu:2000jt, Dermisek:2005ar}.
For example, if $m_h\gg m_{\eta} \gtrsim 10 \GeV$, the dominant decay mode
of $\eta$  is  $\eta\rightarrow b\bar b$  and the experimental lower bound on $m_h$ from $h\rightarrow 4 b$ searches  is around 110 GeV.
This bound can even go down to 86 GeV for $10\GeV\gtrsim m_{\eta} \gtrsim 3.5 \GeV$, where
the dominant  decay mode is   $\eta\rightarrow \tau\bar \tau$ \cite{Chang:2008cw}.

Although technically natural, there is,
{\em a priori}, no reason to believe that all $\epsilon_i$ should be close to $\pm1$, but not exactly $\pm1$ (otherwise $\eta$ is a PQ-axion),
and therefore one might think that  the light-$\eta$ scenario  is not very well motivated.
Nevertheless, it is perhaps reasonable to consider that the
values of  $\epsilon_i$ for the up-type quarks, $\epsilon_u$,  are  different from those of the down-type quark, $\epsilon_d$,  or even from those of the leptons, $\epsilon_l$, and, furthermore,  that  one or more of these are $\pm1$.
In this case we can find natural scenarios in which   $\eta$ is light.
For example, if we assume  $\epsilon_u=\pm 1$ and
$\epsilon_d\not=\pm 1$, we have that $\eta$  receives its mass predominantly
from  a $b_R$ loop, giving
\begin{equation}
\label{masseta}
m^2_\eta\sim \frac{m_b\Lambda_S^3}{16\pi^2 \langle h\rangle f}\simeq \left(30\, {\rm GeV}\right)^2\,
\left(\frac{\Lambda_S}{2\, {\rm TeV}}\right)^3\left(\frac{500\, {\rm GeV}}{f}\right)\, ,
\end{equation}
that is light enough to allow the decay of $h$ to two $\eta$.
The $\eta$ will mainly decay to $b\bar b$, unless  $\epsilon_d=0$. In this latter case, we have that   $\eta$ does not couple to $b\bar b$ and decays
instead to $\tau\bar \tau$. This decay channel can also be zero if  $\epsilon_l=0$, implying that
$\eta$ will mostly decay to $c\bar c$.

Another possibility is to have $\epsilon_u=\epsilon_d=\pm 1$ but $\epsilon_l\not=\pm 1$.
Then the mass of $\eta$ comes from loops of $\tau$ (similar to Eq.~(\ref{masseta}), but with
$m_b\rightarrow m_\tau$), leading to a slightly lighter $\eta$.
In this case, it could be kinematically forbidden for $\eta$ to decay into $b\bar b$,   its principal decay mode then being
into either $c\bar c$ or $\tau\bar \tau$, depending  on whether
$\epsilon_l=0$  or not.

{\bf FCNC:}
Let us now consider the case in which the values of $\epsilon_i$ are not family symmetric.
We expect   FCNC effects mediated at tree-level by $\eta$, which couples linearly to
$\bar f^i_Lf^j_R$  with a strength (assuming $\vev{\eta}=0$ and $\vev{h}\ll 1$)
\begin{equation}
\label{fcnc}
{\cal M}_{ij}=m_{f_i}\, \sum_k U_{R\, ik}\, \epsilon_k\, U_{R\, kj}^\dagger\, ,
\end{equation}
where $U_R$ is the  rotation in  the right-handed sector needed to  diagonalize the
fermion mass matrices and $i,j,k$ runs over all
fermions.
Since $U_R$ is unitary, $U_{R}  U_{R}^\dagger=1$,
we have that, as expected,   $\cal M$ is diagonal for universal values of ${\epsilon_i}$.
We will assume that $U_R$ is of the same order as the CKM matrix  $V$ and study  the implications of non-universality
of $\epsilon_i$ on flavour observables.

In the down-sector, the strongest constraints on FCNC  arise   from  $\Delta m_K/m_K$ and $\varepsilon_K$.
At tree-level, we have that  $\eta$ gives a contribution  to $\Delta m_K/m_K$ given by
\begin{equation}
\label{dmk}
\frac{\Delta m_K}{m_K} = \frac{\text{Re}[{\cal M}_{sd}^2]}{2m^2_\eta f^2m_K}
\langle K|(\bar s_L d_R)^2|\bar K\rangle
\, ,
\end{equation}
where
${\cal M}_{sd}\simeq m_s \{V_{us}V_{ud}[\epsilon_s-\epsilon_d]\}$.
We find  $\Delta m_K/m_K\sim 10^{-15} (100\ {\rm GeV}/m_\eta)^2$,
which is below the experimental bound, $\Delta m_K/m_K\lesssim 7\cdot 10^{-15}$,
for  $m_\eta\gtrsim 40$ GeV.
The bound from  $\varepsilon_K$ can increase the  bound  on the $\eta$ mass by a factor of 10,
but this depends on the phases of $\epsilon_i$ and $U_R$; the constraints from $\Delta m_B/m_B$ are found to be weaker.
Similarly, for the  up sector,    non-universal values for  $\epsilon_i$ lead to
contributions to $\Delta m_D/m_D$. We find that these are  of order $10^{-13}$, and then close to the experimental value,
for   $m_\eta\sim 100$ GeV.
Finally, in the   lepton sector, $\eta$ can induce   contributions to, for example,
$ \tau\rightarrow 3 \mu$, but these are very small and
only  reach   the experimental bound
BR$(\tau\rightarrow 3 \mu)\lesssim 2\cdot 10^{-7}$  for    $\eta$  weighing a  few GeV.

An interesting consequence of having  non-universal values for $\epsilon_i$ is that $\eta$
can   have family-violating decays with a   width  given by
\begin{equation}
\label{fcncdec}
\Gamma(\eta\rightarrow \bar f_i f_j)=\frac{ N_c |{\cal M}_{ij}|^2 m_\eta \beta^4}{8\pi f^2}\, ,\; \; \beta=\sqrt{1- m_i^2/m_\redeta^2}\, ,
\end{equation}
where  $N_c=3$ for quarks and $N_c=1$ for leptons, and we have assumed $m_i\gg m_j$.
If kinematically allowed,
the decay channel $\eta\rightarrow t\bar c$
can be the dominant  one. Indeed, we find
\begin{equation}
\frac{\Gamma(\eta\rightarrow t\bar c)}{\Gamma(\eta\rightarrow b\bar b)}\sim
\frac{|{\cal M}_{tc}|^2}{|{\cal M}_{bb}|^2}\sim \frac{m^2_t V^2_{ts}}{m^2_b} \sim 4\, .
\end{equation}
For a lighter $\eta$,
the  decay channel   $\eta\rightarrow b\bar s$ could     dominate over the $b\bar b$ channel
if  $\epsilon_b=0$,  since in this case one finds
$\Gamma(\eta\rightarrow b\bar s)/\Gamma(\eta\rightarrow b\bar b)\sim
|{\cal M}_{bs}|^2/|{\cal M}_{bb}|^2\sim V^{-2}_{bc}\gg 1$.

\section{Anomalies, the Wess-Zumino-Witten Term, and $CP$}\label{wzw}
Yet another interesting aspect of the phenomenology of models based on the coset $SO(6)/SO(5)$ is that they admit a Wess-Zumino-Witten (WZW) term in their effective lagrangian. As we shall see in more detail below, such terms are interesting for at least three reasons.
First and foremost, the WZW term is the low-energy manifestation of the anomaly
structure of the UV theory (just as  the axial anomaly of the chiral lagrangian in hadronic physics is fixed by the quark content of QCD).
Since it is non-renormalized, it opens a low-energy window onto UV physics:  If it is present in a strongly-coupled theory of EWSB, and if  it is observable at the LHC or a future collider, it would offer a unique opportunity to learn about the UV completion of the theory that controls the weak scale. 
Second, the WZW term gives the leading order correction to the two-derivative sigma model lagrangian, and, 
third, it plays an important  r\^{o}le in the context of discrete symmetries, in particular $CP$.

Before discussing all this in more detail, let us first discuss, in general terms, the conditions for a WZW term to be present in a model of EWSB.
For a sigma model based on the coset $G/H$, there are non-trivial conditions for a WZW term to be present  even when the group $G$ is not gauged.
The condition \cite{AlvarezGaume:1985yb} is that a WZW term, corresponding to an anomalous rep.\ of $G$, can be included only if the anomaly,
restricted to the subgroup $H$, is cancelled by the $H$ anomaly of massless fermions present in the low-energy effective theory.
To see why the anomaly must match in this way, we note that the sigma model has a local $H$ symmetry, corresponding to a compensating transformation that maintains the parametrization of the coset $G/H$; it is local because the coset parametrization is written in terms of the spacetime-dependent NGB fields.
To maintain the Ward identities, which, in particular, guarantee that NGBs are massless, $H$ must be anomaly free.

If we wish to go further and gauge all of $G$, then the $H$-anomaly of light fermions must itself vanish \cite{Preskill:1990fr}. If it does not, then by sandwiching together two triangle diagrams involving the light fermions and three $H$ gauge bosons, we can generate masses for the $H$ gauge bosons, and these cannot be cancelled by diagrams involving a WZW term and NGBs.
Then the argument of the previous paragraph tells us that the $H$-anomaly of any WZW term must vanish.

For theories of EWSB, we do not gauge all of $G$, but only some subgroup $K$, which intersects non-trivially with the unbroken group $H$.
As a result, the surviving massless gauge fields belong not to $K$ or $H$, but rather to some smaller subgroup $J$ that is common to both $K$ and $H$.
In this general case, we can only apply the logic of the previous argument to the group $J$.

So in summary, a necessary condition for a WZW term is that $G$ admits anomalous representations whose anomalies vanish when construed as anomalies of
the surviving linearly-realized gauge symmetry $J \subset H,K$.

Now let us consider the implications for some specific examples. A Higgsless model with coset structure $SU(2)_L \times U(1)_Y/U(1)_Q$ satisfies the condition for a WZW term. Indeed, a suitable anomalous rep.\ of $G$ is $(2,-\frac{1}{2}\sqrt{\frac{1}{2}}) \oplus (1,\sqrt{\frac{1}{2}}) $.
For a model with a Higgs and a WZW term, one may consider the coset $SU(3)\times U(1)_X / SU(2)_L \times U(1)_Y $ of Ref.~\cite{Contino:2003ve}. However, the $SO(5)/SO(4)$ model with a Higgs and custodial symmetry does not admit a WZW term, the reason being that $SO(5)$ does not have anomalous representations.\footnote{Identical conclusions were reached from a different direction in Ref.~\cite{Gripaios:2008ei}.}

For an example which incorporates custodial symmetry and can have a WZW term, we need to look no further than the model based on $SO(6) / SO(5)$ that we discussed in the previous Section. Since $SO(6)$ is locally isomorphic to $SU(4)$, it has anomalous representations. Other examples are cosets based on
$SO(6)/ SO(4) $, which we discuss in Appendix~\ref{4},  and models based on $SU(4) \times SU(4)/SU(4)$.

These results are confirmed by consideration of the effective lagrangian. The full form of the WZW term is somewhat complicated, involving an infinite series of terms in the PNGBs; derivations in the context of holographic Higgs models are given in Ref.~\cite{Gripaios:2008ei} and in the context of little Higgs models in Ref.~\cite{Hill:2007nz} (some phenomenological aspects of anomalies in little Higgs models were discussed in Ref.~\cite{Kilian:2004pp}). 
Nevertheless,  at leading  order in $1/f$, 
the WZW term gives  a coupling of a   PNGB to two gauge bosons via the epsilon tensor.
In a Higgsless model, for example, the effective lagrangian should respect the $U(1)_Q$ of electromagnetism, and indeed we can couple the charge neutral PNGB that is eaten by the $Z$ to the electromagnetic field combination $F \tilde{F}$. By contrast, the effective lagrangian in a model with a Higgs should respect the full $SU(2)_L\times U(1)_Y$; no 
operator with  a single  PNGB and two gauge bosons  is available in a theory with just a SM Higgs (like the $SO(5)/SO(4)$ model), but once we add a singlet (for example in an $SO(6)/SO(5)$ or $SU(3)\times U(1)_X / SU(2)_L \times U(1)_Y $ model), we can write down a gauge-invariant term  of the form
\begin{gather}
\label{anom}
\mathcal{L} \subset \frac{\eta} {16\pi^2 } (n_B B_{\mu\nu}\tilde{B}^{\mu\nu} + n_W W_{a\, \mu\nu}\tilde{W}^{a\, \mu\nu} + n_G G_{A\, \mu\nu} \tilde{G}^{A\, \mu\nu}).
\end{gather}
Here, $G_{A\, \mu\nu}$, $W_{a\, \mu\nu}$ and $B_{\mu\nu}$ refer to the field strengths of the $SU(3)_c \times SU(2)_L\times U(1)_Y$   gauge group,  and $\tilde B^{\mu\nu}=\epsilon^{\mu\nu\rho\sigma}B_{\rho\sigma}/2$ and similarly for the other fields.
The $n_{G,W,B}$ are integers that measure the strengths of the various anomalies and are fixed by the fermion content of the UV physics. If we were able to measure these integers at the LHC, then we would gain quantitative information about the UV physics, just as measurement of the decay rate $\pi^0 \rightarrow 2\gamma$ in hadronic physics tells us that the number of colours in QCD is three. Note that for the $SO(6)/SO(5)$ model, the WZW terms all come from an $SU(4)^3$ anomaly, such that $n_G = 0$ and $n_W = n_B$.
However, the SM fermions also give contributions to the terms in Eq.~(\ref{anom}).
Indeed, we find, in the approximation $m_{f_i}\gg m_\eta$:
$\delta n_B=N_c Y^2_i \text{Re}[\epsilon_i] $,  $\delta n_W=N_c \text{Re}[\epsilon_i]$ for weak doublets, and
$\delta n_G=\text{Re}[\epsilon_i]$ for quarks. Since the relevant couplings do not respect the $G$ symmetry, the shifts in the coefficients are not restricted to integers.

But can we measure these coefficients at the LHC, or if not, at a future collider?
If we cannot measure the coefficients themselves, can we even detect the presence of these terms? We might hope to be able to produce the  $\eta$ directly at the LHC via  $WW$ fusion and the $\eta W \tilde{W}$ vertex, or via gluon fusion if an $\eta G \tilde{G}$ vertex is present.
Alternatively, and similar to the neutral pion in QCD,
we could measure  the coefficients by detecting
the  decay of $\eta$ to photons. We find
\begin{gather}
\frac{\Gamma(\redeta\rightarrow \gamma\gamma)}{\Gamma(\eta\rightarrow b\bar b)}
\simeq 0.007\, \Big|\frac{n_\gamma}{5}\Big|^2\Big(\frac{m_\eta}{100\ {\rm GeV}}\Big)^2\left|\frac{1}{\epsilon_b}\right|^2\, ,
\end{gather}
where $n_\gamma=n_B+n_W$.
Although this partial width is small, it is larger, for $n_\gamma\sim 5$,  than
the SM decay $\Gamma(h\rightarrow \gamma\gamma)$
which has been shown to  be visible at the LHC.
Furthermore, the branching ratio of  $\eta$ to photons
can be enhanced if, as we explained in the previous section,
the $\eta$ cannot decay to $b\bar b$.
We must however  stress that even if we are able to observe the decay channel
to photons, at the LHC we can only measure  the product of the cross section and the branching ratio, not the partial width. So we cannot extract the strength of the anomaly directly, without further information. A final possibility is that once we include the higher mass resonances
in the effective theory, we expect that they too will couple via the anomaly (an extra-dimensional example was recently discussed in Ref.~\cite{Csaki:2009re}),
giving us another potential experimental window on the couplings.

Let us lastly discuss the connection between the WZW term and discrete symmetries, in particular $CP$.
Discrete symmetries were, of course,
the very reason for the introduction of the WZW term in the chiral lagrangian of QCD, at least in Witten's incarnation thereof \cite{Witten:1983tw}. To recall, the leading order (two-derivative)
chiral lagrangian, $\tr (\partial e^{i\pi} \partial e^{-i\pi})$ is invariant under the na\"{\i}ve parity, $P_0: x \rightarrow -x$, as well as the NGB parity,  $P_{NGB}: \pi \rightarrow - \pi$,
and charge conjugation, $C: \pi \rightarrow \pi^T$. However, of the first two, only the true parity $P = P_0 P_{NGB}$ is a symmetry of QCD, and the WZW term is the leading order term in the chiral lagrangian that violates $P_0$ and $P_{NGB}$ individually, while respecting $P$.
In Appendix \ref{CP} we show that analogous arguments go through for the EWSB model based on the coset $SO(6)/SO(5)$:
The lagrangian for the gauge and Higgs self-interactions, including the WZW term,
respects $CP$ if $h$ and $\eta$ are defined to be $CP$-even and $CP$-odd respectively.

\section{Outlook}\label{conclu}
We have explored a composite Higgs model based on the coset $SO(6)/SO(5)$, with SM fermions assigned to the $\mathbf{6}$ of $SO(6)$.
Just like the minimal composite model based on $SO(5)/SO(4)$, the model features custodial protection of the $T$-parameter and $Z \rightarrow b \overline{b}$, and therefore is in agreement
with EWPT if a mild tuning of $v/f$ is accepted to accommodate the $S$-parameter.\footnote{In this model the contribution to $S$ is equal to that in the minimal composite Higgs model \cite{Agashe:2004rs}.}
The model contains an extra singlet scalar, $\eta$, compared to the Higgs sector of the SM, which can dramatically change the phenomenology.
This strongly  depends on the values of
$\epsilon_i$ that, as can be seen from Eq.~(\ref{specialv}),
determines the  properties of $\eta$.
In particular, we have presented scenarios
in which  the SM Higgs can predominantly decay into $2\eta$, which in turn can dominantly decay  into any one of $b\overline{b}$, $\tau \overline{\tau}$, or
$c\overline{c}$.
As a result, the direct bound on the SM Higgs mass coming from LEP can be invalidated, and the true bound may in fact be as low as 86 GeV.
The couplings of the singlet to SM fermions can also give rise to tree-level FCNCs that are
close to  the  experimental bounds (or even exceeding  it in the case of  $\varepsilon_K$), and induce flavour-violating decays for $\eta$.
The model can also exhibit explicit or spontaneous  $CP$ violation, though
it will be difficult to test experimentally.
One of the most interesting phenomenological aspects of the model
is the coupling of $\eta$ to gauge bosons, which is induced not only by SM loops,
but also can be present if the model has anomalies.
Therefore the process $\eta\rightarrow\gamma\gamma$
will be  of crucial importance to unravel the underlying structure of the model.

At the LHC
the most prominent  way to produce  the $\eta$ is either through  the decay of the Higgs
$gg\rightarrow h\rightarrow \eta\eta$, if kinematically allowed,  or from gluon fusion $gg\rightarrow\eta$.
Nevertheless, $\eta$ can also be produced in the decay of a heavy fermionic resonance
of the strong sector.
Its detection is, however,  difficult. The most promising
decay channel   is $\eta\rightarrow\gamma\gamma$,
which we expect to have a partial width  larger   than that of  the  corresponding SM Higgs decay.
The phenomenological prospect   at the LHC and other future colliders
need, however,  to be fully explored.

The model presented here can also have  interesting implications for astrophysics. For example, if  $\epsilon_i=0$ for all SM fermions   and there are no anomalies, the singlet
is stable, and hence can be a dark matter candidate. The singlet can annihilate through the $h^2 \eta^2$ interactions of Eqs.~(\ref{kin}) and (\ref{pot}) and these determine the relic density.
The resulting physics is presumably not dissimilar from that discussed in Ref.~\cite{McDonald:1993ex,McDonald:1993ey,Davoudiasl:2004be}.
Another interesting question is  whether electroweak baryogenesis can be realized in the model.
The SM fails in this regard, because the $CP$ violation in the CKM matrix is too small and the electroweak phase transition cannot be strongly first-order
given the LEP bound on the Higgs mass. In the $SO(6)/SO(5)$ model, the presence of the singlet could cure both of these problems.
First, we have shown that, for  $\langle \eta\rangle\not=0$,  the model has  new sources of  $CP$ violation. Secondly,  the presence of the singlet, as shown in Ref.~\cite{Anderson:1991zb}, can result in a strongly first-order phase transition for Higgs masses above the LEP bound.
All these issues deserve further analysis.

\begin{acknowledgments}
BMG thanks R.~Rattazzi for discussions.
FR was supported by the European Community's Research Training Network (MEST-CT-2005-020238-EUROTHEPHY) and the Swiss National Fund.
The  work  of AP and JS was  partly supported   by the
Research Projects CICYT-FEDER-FPA2005-02211,
SGR2005-00916,  ?UniverseNet? (MRTN-CT-2006-035863),
and  AP2006-03102.

\end{acknowledgments}
\appendix
\section{Models based on $SO(6)/SO(4)$}\label{4}

In the case in which the breaking of the $SU(4)$ is achieved by the VEV of the  symmetric  representation, the ${\bf 10}$, the global $SU(4)$ is broken down to $SO(4)$.
In this case, however, the nine  NGBs parametrizing $SU(4)/SO(4)$ transform as a ${\bf (3,3)}$
of $SO(4) \cong SU(2)_L \times SU(2)_R$,  which does not contain doublets that can be associated with the SM Higgs.

Another option is to break $SU(4)$  by the VEV of the traceless representation, the ${\bf 15}$, that
we denote by $\Omega$ and transforms as $\Omega\rightarrow U\Omega U^\dagger$.
When the VEV of $\Omega$ takes the form
\begin{gather}
\label{vevo}
\Omega_0=\mathrm{Diag}(1,1,-1,-1) \, ,
\end{gather}
the global  $SU(4)$ is broken down to   $SU(2)_L \times SU(2)_R\times U(1)\cong SO(4)\times SO(2)$,  delivering 8 NGBs, which transform
as $(\mathbf{2},\mathbf{2})_{\pm 2}$ under the unbroken subgroup.
This  model has two Higgs doublets, which gives rise to the following  problem.
While a  single Higgs doublet automatically guarantees that, after EWSB,
the  global $SO(4)$ symmetry of the strong sector is broken down to  the custodial $SO(3)$ symmetry that  protects the $T$-parameter from receiving large corrections,
the presence of two Higgs doublets  spoils this property.
The reason is that the second Higgs doublet can get a VEV, breaking the custodial $SO(3)$ symmetry down to $SO(2)$.
To see this explicitly,
let us parametrize the NGBs by  the traceless, hermitian matrix
\begin{gather}\label{PIOMEGA}
\Omega=e^{\frac{1}{\sqrt{2}}i\Pi_{\Omega}/f}\Omega_0\, ,\ \ \ \
\Pi_{\Omega}=\begin{pmatrix} 0 &\hat{H_1}+i \hat{H_2} \\ \hat H_1^\dagger-i \hat H_2^\dagger&0 \end{pmatrix},
\end{gather}
where $\hat H_i=(H^c_i, H_i)$. By an $SU(2)_L$ rotation, we can eliminate 3 out of the 4 components of $\hat H_1$ and write $\hat H_1=h\mathds{1}$. For $\hat H_2$, we
only consider the $SO(3)$-breaking direction  $\hat H_2=-ih_3\sigma_3$.
For simplicity,  we will take the limit $h,h_3\ll f$, which allows us to expand Eq.~(\ref{PIOMEGA}):
\begin{equation}
\Omega\simeq \begin{pmatrix} (1-\frac{h^2+h^2_3}{4f^2})\mathds{1}-\frac{h h_3}{2f^2}\sigma_3 &-\frac{i}{f\sqrt{2}}(h\mathds{1}+h_3\sigma_3) \\
\frac{i}{f\sqrt{2}}(h\mathds{1}+h_3\sigma_3)&-(1-\frac{h^2+h^2_3}{4f^2})\mathds{1}+\frac{h h_3}{2f^2}\sigma_3 \end{pmatrix}.
\end{equation}
From the kinetic term of $\Omega$ we can read off the SM gauge boson masses:
\begin{equation}
\label{kin2}
\frac{f^2}{8}\tr |D_{\mu} \Omega|^2 =\frac{g^2}{8}{(h^2+h^2_3)}\left[ W^{\mu+} W^-_\mu+\frac{1}{2\cos^2\theta_W}\left(1-\frac{h^2h^2_3}{2f^2(h^2+h^2_3)}\right)Z^\mu Z_\mu\right]+\cdots\, ,
\end{equation}
which shows that if $h_3$ gets a VEV, the custodial symmetry is broken and $\rho\equiv m^2_W/(m_Z^2\cos^2\theta_W)\not= 1$.
Now, let us choose that the SM top be embedded in a  $\bf 6$ of $SU(4)$, as in Eq.~(\ref{embe1}) (similar results are obtained for the ${\bf 10}$ representation).
This implies that the strong sector generates the operator
\begin{equation}
\tr[\bar\Psi_q \pslash\Psi_q \Omega^*]=-\bar u_L \pslash u_L \frac{h h_3}{8f^2}+\cdots\, .
\end{equation}
This coupling can enter in a $u_L$-loop and generate (after EWSB $\vev{h}\not=0$)  a tadpole for $h_3$; this   forces $h_3$  to  get a  VEV, breaking   the custodial symmetry.
This poses a serious problem for this type of model.

Finally, we can consider   the global symmetry breaking $SU(4)\rightarrow SO(4)$
achieved by the presence of the  VEV of $\Omega$ -Eq.~(\ref{vevo})- and $\Sigma$ -Eq.~(\ref{SigmaVEV}).
In this case there are 9 NGBs
transforming  as $(\mathbf{1},\mathbf{1}) \oplus (\mathbf{2},\mathbf{2})\oplus (\mathbf{2},\mathbf{2})$.
These models, however, not only suffer from the problems discussed above
but   can also have sizable   FCNC, since the Yukawa couplings can arise
from two distinct multiplets, $\Sigma$ and $\Omega$.

\section{$CP$-invariance}\label{CP}

To show that $CP$ is a symmetry of the sigma model representing the Higgs sector of the $SO(6)/SO(5)$ model, we begin by asserting that the Lie algebra of $SO(6)$ admits two automorphisms, given by
\begin{align} \label{autos}
A_1: \; T^a  &\rightarrow T^a, T^{\hat{a}} \rightarrow - T^{\hat{a}}\, , \nonumber \\
A_2: \;T^a &\rightarrow -T^{aT}, T^{\hat{a}} \rightarrow  T^{\hat{a}T}\, ,
\end{align}
where $T^a$ are the generators of the unbroken $SO(5)$ and $T^{\hat{a}}$ are the broken generators, as in Eqs.~(\ref{unbroken}) and (\ref{broken}). Recall that an automorphism
is a linear transformation among the generators that preserves the algebra.  That the two transformations in Eq.~(\ref{autos}) preserve the algebra follows from the fact that
$SO(6)/SO(5)$ is a symmetric space: There exists a basis for $\mathrm{Lie} \; SO(6) $ (such as the explicit one given after Eqs.~(\ref{unbroken}) and (\ref{broken})) such that
\begin{equation} \label{basis}
[T^a,T^b]   \sim  T^c\, , \, \, \,
[T^{\hat{a}}, T^{\hat{a}}] \sim T^a\, , \, \, \,
[T^{\hat{a}}, T^{a}] \sim T^{\hat{b}}\, .
\end{equation}
It remains only to check that $A_2$, which involves transposition, can be written as a linear transformation among the generators. This is easily done
using the explicit representation for the generators given after Eqs.~(\ref{unbroken}) and (\ref{broken}). Note also that since our sigma model field is written as
an exponential of the broken generators, $\Sigma \sim e^{i \Pi_{\hat{a}} T^{\hat{a}} }$, these automorphisms can also be thought of as the field
transformations $\Pi \rightarrow - \Pi$ and $\Pi \rightarrow  \Pi^T$,
just as in the chiral lagrangian for QCD.

How do these two automorphisms give rise to symmetries of the sigma model lagrangian? To answer this, we note that the
general $G/H$ coset sigma model is constructed in the following way. Firstly, given a coset representative $\Sigma$ for $G/H$, we build the Cartan form for $G$, $\Sigma^{-1} d \Sigma$, which is of course an element of
$\mathrm{Lie} \; G $. Projecting this onto the subspace of broken generators, $(\Sigma^{-1} d \Sigma)_{\hat{a}}$ gives a vielbein corresponding to the natural metric on $G/H$. The vielbein is the basic object that we use to build the sigma model. In particular, the leading two-derivative term in the sigma-model lagrangian is
just the natural metric on $G/H$, built out of two vielbeine, and pulled back to spacetime. Similarly, the WZW term (in $d=4$) is built out of five vielbeine.
Now the automorphisms give rise to isometries of the natural $G/H$ metric, so any terms in the sigma-model lagrangian built out of the metric will be symmetric. The WZW term is special in that it is built not out of the metric {\em per se}, but out of the vielbein. Under the automorphism $A_1$ in Eq.~(\ref{autos}),
the vielbein changes sign, and the WZW term also changes sign. So $A_1$ is not a symmetry of the WZW term. However, when combined with the spacetime parity operation, $P_0:\; x\rightarrow -x$, the WZW term (which features four derivatives and an epsilon tensor)  is invariant.

So we have proven that for a general $G/H$ symmetric space, with the two automorphisms in Eq.~(\ref{autos})~\footnote{With the caveat that $A_2$ be a {\em bona fide} automorphism.} the sigma model lagrangian, including the WZW term, will be invariant under two symmetries, corresponding to $A_2$ and $A_1 P_0$.
In the $SO(6)/SO(5)$ model of EWSB, the combination $A_1 A_2 P_0$ corresponds precisely
to
\begin{equation}
h\rightarrow h\ , \ \ \ \  \eta\rightarrow -\eta\, .
\end{equation}
This  defines  the  $CP$  symmetry of the Higgs sector, including the WZW term.
This is, of course, in accord with  the WZW contribution Eq.~(\ref{anom})
which couples the $CP$-odd $\eta$ to the $CP$-odd combination $F\tilde{F}$.

\bibliography{nmchmrefs}
\end{document}